\documentclass [12pt,twoside]{article}           % version LATEX2E
\usepackage[left,modulo]{lineno}
\usepackage{setspace}
\countdef\arxiv=201
\ifnum\arxiv=0 \input cpreb.tex \fi

\ifnum\arxiv=1

\usepackage{epsfig,times,lscape}
\usepackage[usenames]{color}

    \ifnum\count102=1

    \fi

    \ifnum\count102=2
\fi

\newcommand{\nc}{\newcommand}

     \ifnum\count101=1
\nc{\qI}[1]{\section{{#1}}}
\nc{\qA}[1]{\subsection{{#1}}}
\nc{\qun}[1]{\subsubsection{{#1}}}
\nc{\qa}[1]{\paragraph{{#1}}}

\def\qpar{\vskip 2mm plus 0.2mm minus 0.2mm}
\def\qL{\hfill \break}
     \fi 

      \ifnum\count101=2
 \nc{\qI}[1]{\parindent=0mm \vskip 8mm 
{\centerline{\LARGE \color{red}#1}}\vskip 3mm}
\nc{\qA}[1]{\vskip 2.5mm \noindent 
{{\bf\large\color{blue}  #1}} \vskip 1mm \parindent=0mm}
 \nc{\qun}[1]{\vskip 1mm \noindent {\sl #1 }\quad }

\def\qL{\hfill \break}
\def\qpar{\vskip 2mm plus 0.2mm minus 0.2mm}

      \fi
\def\qth{\vrule height 12pt depth 0pt width 0pt}
\def\qtb{\vrule height 0pt depth 5pt width 0pt}

\nc{\qfoot}[1]{\footnote{{#1}}}

      \ifnum\count101=1
\def\qbu{\hfill \par \hskip 6mm $ \bullet $ \hskip 2mm}
\def\qee#1{\hfill \par \hskip 6mm (#1) \hskip 2 mm}
      \fi
      \ifnum\count101=2
\def\qbu{\hfill \par \hskip 4mm $ \bullet $ \hskip 2mm}
\def\qee#1{\hfill \par \hskip 4mm (#1) \hskip 2 mm}
      \fi

\def\qparr{ \vskip 1.0mm plus 0.2mm minus 0.2mm \hangindent=10mm
\hangafter=1}

     \ifnum\count101=1 
 
     \fi
     \ifnum\count101=2

  \def\qcitb#1{\noindent \hbox to 102mm{\hfill \small #1} \vskip 1mm}
      \fi

 \def\qpages#1{\count102=0{\loop\advance\count102 by 1
 \null \vfill\eject \ifnum\count102<#1 \repeat}}

\def\qth{\vrule height 12pt depth 0pt width 0pt}
\def\qtb{\vrule height 0pt depth 5pt width 0pt}

\def\qv{\vskip 0.1mm plus 0.05mm minus 0.05mm}
\def\qhu{\hskip 0.6mm}
\def\qhv{\hskip 3mm}

\def\qhw{\hskip 1.5mm}
\def\qleg#1#2#3{\noindent {\bf \small #1\qhw}{\small #2\qhw}{\it \small #3}\qv }
\newcommand{\promille}{%
  \relax\ifmmode\promillezeichen
        \else\leavevmode\(\mathsurround=0pt\promillezeichen\)\fi}
\newcommand{\promillezeichen}{%
  \kern-.05em%
  \raise.5ex\hbox{\the\scriptfont0 0}%
  \kern-.15em/\kern-.15em%
  \lower.25ex\hbox{\the\scriptfont0 00}}

\fi
\begin{document}
\thispagestyle{empty}

% --------------------------------------------------------------------

      % Hauts de pages et numerotation

          % Remarque: sans le \protect --> message d'erreur (ordre fragile)
\markboth{{\sl \hfill  \hfill \protect\phantom{3}}}
        {{\protect\phantom{3}\sl \hfill  \hfill}}

% -------------------------------------------------------------------
\color{yellow} 
%\hrule height 20mm depth 20mm width 170mm 
\hrule height 10mm depth 10mm width 170mm 
\color{black}

%\vskip -12mm   % pour un titre avec 1 seule ligne
 \vskip -17mm   % pour un titre avec seconde ligne

\centerline{\bf \Large The influence of Lent on marriages and conceptions}
\vskip 5mm
\centerline{\bf \Large explored through a new methodology}
\vskip 10mm

\centerline{\large 
Claudiu Herteliu$ ^1 $, Peter Richmond$ ^2 $ and Bertrand M. Roehner$ ^3 $
}

\vskip 10mm
\large

%{\bf \color{red} SUMMARY}\qL
%{\bf \color{blue} Background:} \quad
%{\bf \color{blue} Aim:} \quad 
%{\bf \color{blue} Method:} \quad 
%{\bf \color{blue} Findings:}\quad 
%{\bf \color{blue} Conclusion:}\quad 
%
{\bf Abstract}

Herteliu et al. (2015) have elsewhere analyzed the impact of religious
festivals on births in Romania. 
Here we broaden the analysis (i) by studying the effect of
Lent on marriages as well as births (ii) by analyzing
a number of other countries which allows a comparison with
non-Orthodox countries.
We also introduce a new methodology which
treats the data in a way that frees the analysis from bias
related to seasonal patterns of births and marriages. \qL
The comparison between the effects on
marriages and conceptions appears
of particular interest for it permits to assess the respective
weighs of social
pressure on one hand and personal leanings on the 
other hand.
Our analysis reveals a strong effect
of Lent on marriages with a reduction by 80\% in Orthodox countries 
and 40\% in West European
Catholic and Protestant countries.
Since the influence of Lent on conceptions
is independent of any form of direct social
control one might expect the effect to be much smaller. 
In percentage terms it is
roughly 10 times smaller than the effect on marriage. \qL
The present methodology opens the
way to the accurate investigation of the impact of other
mobile religious periods (e.g. Ramadan) on various social phenomena
(e.g. suicide).

% RESUME INITIAL (20 mars 2018)
\count101=0  \ifnum\count101=1
With respect to a previous paper by Herteliu et al. (2015)
of which it is a continuation, the present paper contains
two innovations. 
(i) It gives a comparative view across different countries. 
(ii) It introduces
a new methodology in which successive months are treated
separately which frees the analysis from any bias
related to seasonal patterns of births or marriages.\qL
When applied to the case of marriages, the analysis
reveals a very strong incidence of Lent, with a division
by about 5 in Orthodox countries and a reduction by about
40\% in west-European countries. \qL
The incidence of
Lent on conceptions is of a different kind in the sense
that it is independent of any form of social control.
Therefore, one expects it to be much smaller.
In terms of percentage the reduction is 
roughly ten times smaller than for marriages. 
Using the same methodology it will in the future be possible
to extend this exploration to the incidence of special
religious periods
on other vital rates, for instance suicide rates. 
\fi

\vskip 10mm
\centerline{\it \small Version of 7 April 2018}
\centerline{\it \small Provisional. Comments are welcome.}
\vskip 5mm

{\small Key-words: religion, birth, marriage, Lent, Ramadan} 

\vskip 5mm

{\normalsize
1: Department of Statistics and Econometrics, Bucharest University
of Economic Studies, Bucharest, Romania. 
Email: claudiu.herteliu@gmail.com \qL
2: School of Physics, Trinity College Dublin, Ireland.
Email: peter\_richmond@ymail.com \qL
3: Institute for Theoretical and High Energy Physics (LPTHE),
University Pierre and Marie Curie, Sorbonne Universit\'e,
Centre de la Recherche Scientifique (CNRS).
Paris, France. \qL
Email: roehner@lpthe.jussieu.fr
}

\vfill\eject

\qI{Introduction}

The purpose of our investigation is to find explanations
for some puzzling observations in relation
with marriage and birth rates.
Although we know that nowadays it is fairly uncommon to
start a paper by describing intriguing facts%
\qfoot{In the 18th and 19th centuries, it was common for
Academies of 
Sciences to organize academic contests (rewarded by a 
cash prize) in which the participants had to
explain some challenging facts. For instance,
one of those competitions
which took place in France in 1818, led to the discovery
of the so-called Arago spot, a bright dot at the center of 
the shadow of an opaque circular object. This unexpected
and astounding observation became a key argument in favor of the
wave theory of light.}%
, 
in the present
case we felt that this presentation could be useful
because, beyond the specific issues treated here, it will also
emphasize that comparative analysis can
be of great help in solving such riddles.

\qA{Marriage rates in March}

Our first intriguing observation is the strange pattern of
marriage numbers during the month of March in Bulgaria.
Such a more or less periodic plot looks like
seasonal fluctuations except that here the curve shows
{\it annual} variations in a given month. Why should the
number of marriages in 1929 be over 10 times larger than
in 1928? 
\qpar

%
%%-----------------------------------------------
%%%%   MARIAGES DU MOIS DE MARS EN BULGARIE ET AU JAPON 
\begin{figure}[htb]
\centerline{\psfig{width=11cm,figure=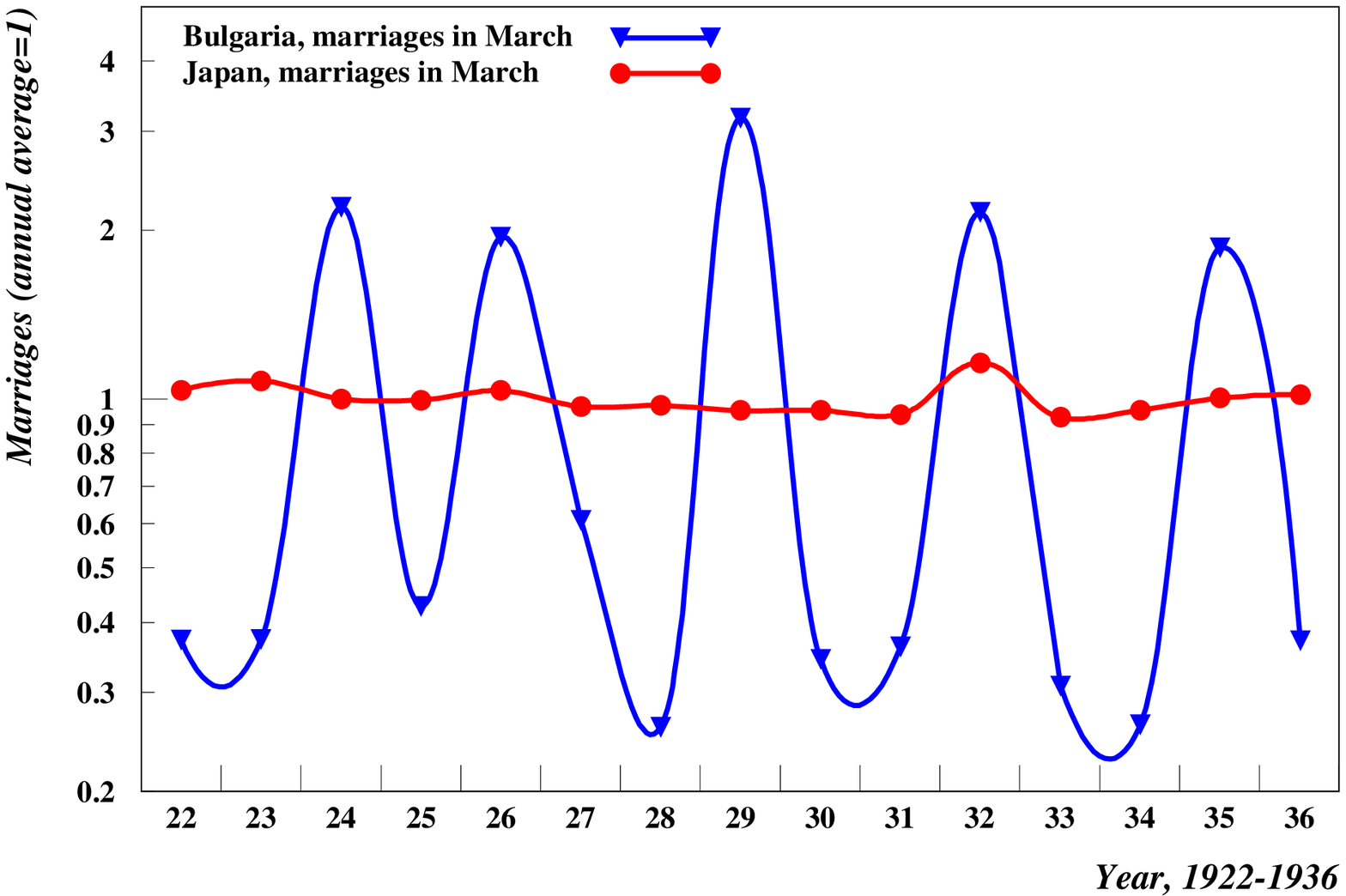}}
\qleg{Fig.\qhu 1a\qhv Number of marriages during March
in Bulgaria and Japan, 1922-1936.}
{For the purpose of comparison the two series were
normalized. Note the broad range (from 1 to 10) of 
the Bulgarian changes.
Clearly there is a factor at work in Bulgaria
which does not exist in Japan.}
{Source: Bunle (1954, p. 249 and 258.}
\end{figure}
%-------------------------------------------------
%
%%-----------------------------------------------
%%%%   MARIAGES DU MOIS DE MARS: BULGARIE,FRANCE,ROMANIA,GREECE
\begin{figure}[htb]
\centerline{\psfig{width=11cm,figure=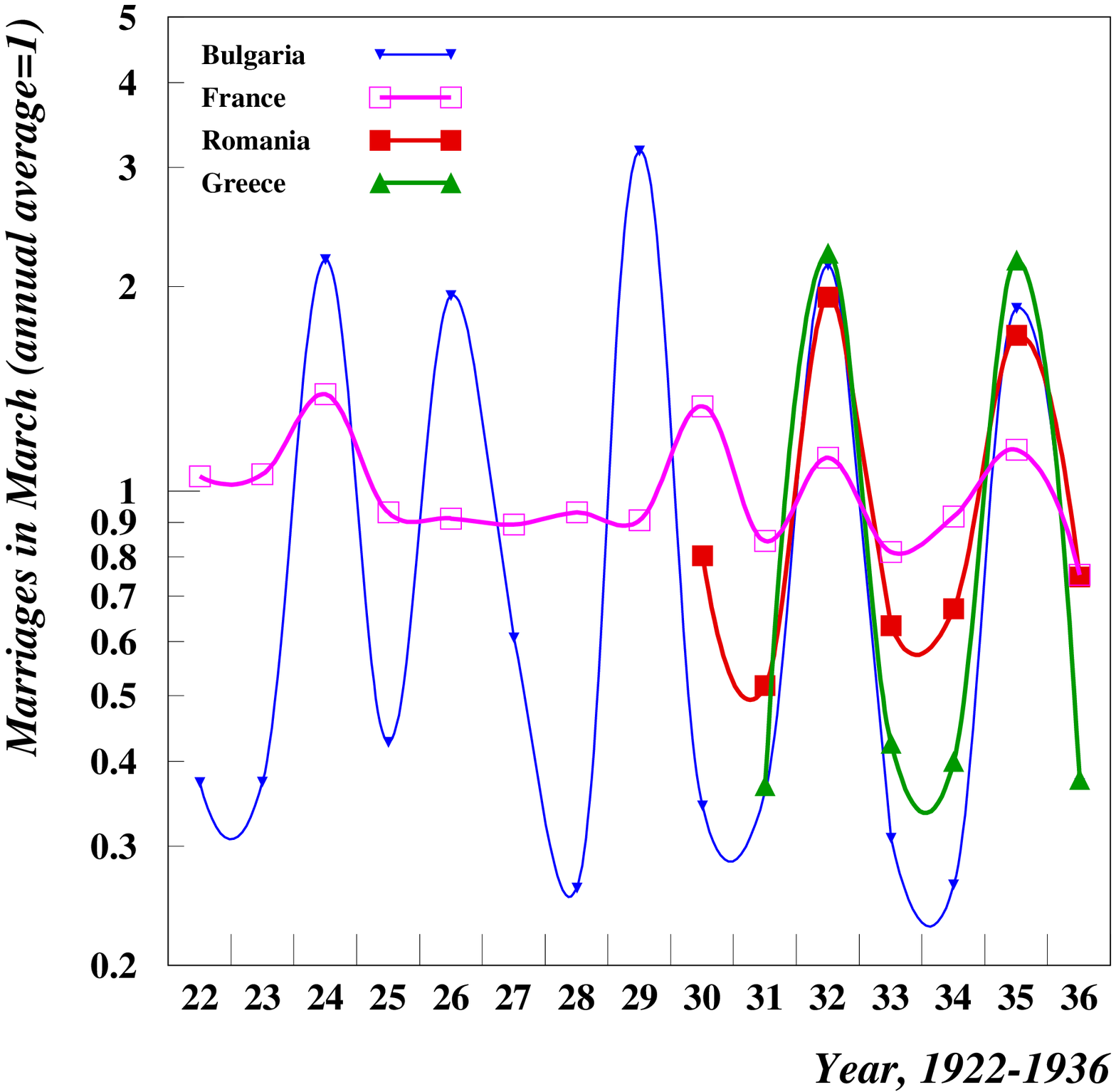}}
\qleg{Fig.\qhu 1b\qhv Number of marriages during March
in Bulgaria, France, Romania and Greece.}
{For Romania and Greece the data are available only from
1930 and 1931 on. For France the fluctuations are somewhat
similar to those in Bulgaria yet not as identical as those
of Romania and Greece.}
{Source: Bunle (1954, p. 249-251 and 255.}
\end{figure}
%-------------------------------------------------

Can comparisons with other countries help us to
identify the origin of this mysterious pattern?
\qpar

As a first test, it is natural to try a distant country 
which differs from Bulgaria in many respects. Fig. 1a
shows that what we see in Bulgaria is not a ``universal''
regularity but rather one that is tied to
a specific factor which, obviously, does not exist in 
Japan. 
\qpar

As a second test let us consider another European country,
for instance France. The French marriage numbers display
fluctuations which are somewhat similar to those in
Bulgaria although of much smaller amplitude and not
exactly in sync (Fig. 1b). This observation makes us suspect that
in France there is a factor similar, yet not identical,
to the one at work in Bulgaria.
\qpar

Greece and Romania have a common border with Bulgaria
in the south and north respectively. Although their curves
cover only the last years of the interval, the fact that
they are highly parallel strongly suggests the existence
of the same factor as in Bulgaria. The three countries belong
to the Orthodox world. Is this the factor which can explain
the observed fluctuations?

\qparr
A possible mechanism comes to mind.\qL
In the Christian religion the time preceding the Easter
Sunday is called Lent. In the Orthodox religion Lent
lasts 7 weeks (i.e. 49 days). As Lent is a time of fasting
and penance it is understandable that people will avoid
celebrating their marriage during Lent.
Can this mechanism account for the fluctuations observed in
the number of marriages?
\qpar

In order to get a quick answer
let us observe two extreme cases, namely 1928 (613 marriages)
and 1929 (7,462 marriages). In 1928, Orthodox Lent started
on 27 February and ended on 14 April which means that the
whole month of March was included in the Lent period.
On the contrary, in 1929, Lent started on 18 March
which means that only 13 days were included in Lent.
Actually, in the whole series of years from 1922 to 1936 this is
the latest date which explains that 1929 has the largest 
number of marriages.
\qpar

By repeating this procedure (which is summarized in Fig, 2)
for all the years we get the curves shown in Fig. 3.
%
%%-----------------------------------------------
%%%% MISE EN // DES INTERSECTIONS ET DES MARRIAGES: PROCEDURE
\begin{figure}[htb]
\centerline{\psfig{width=11cm,figure=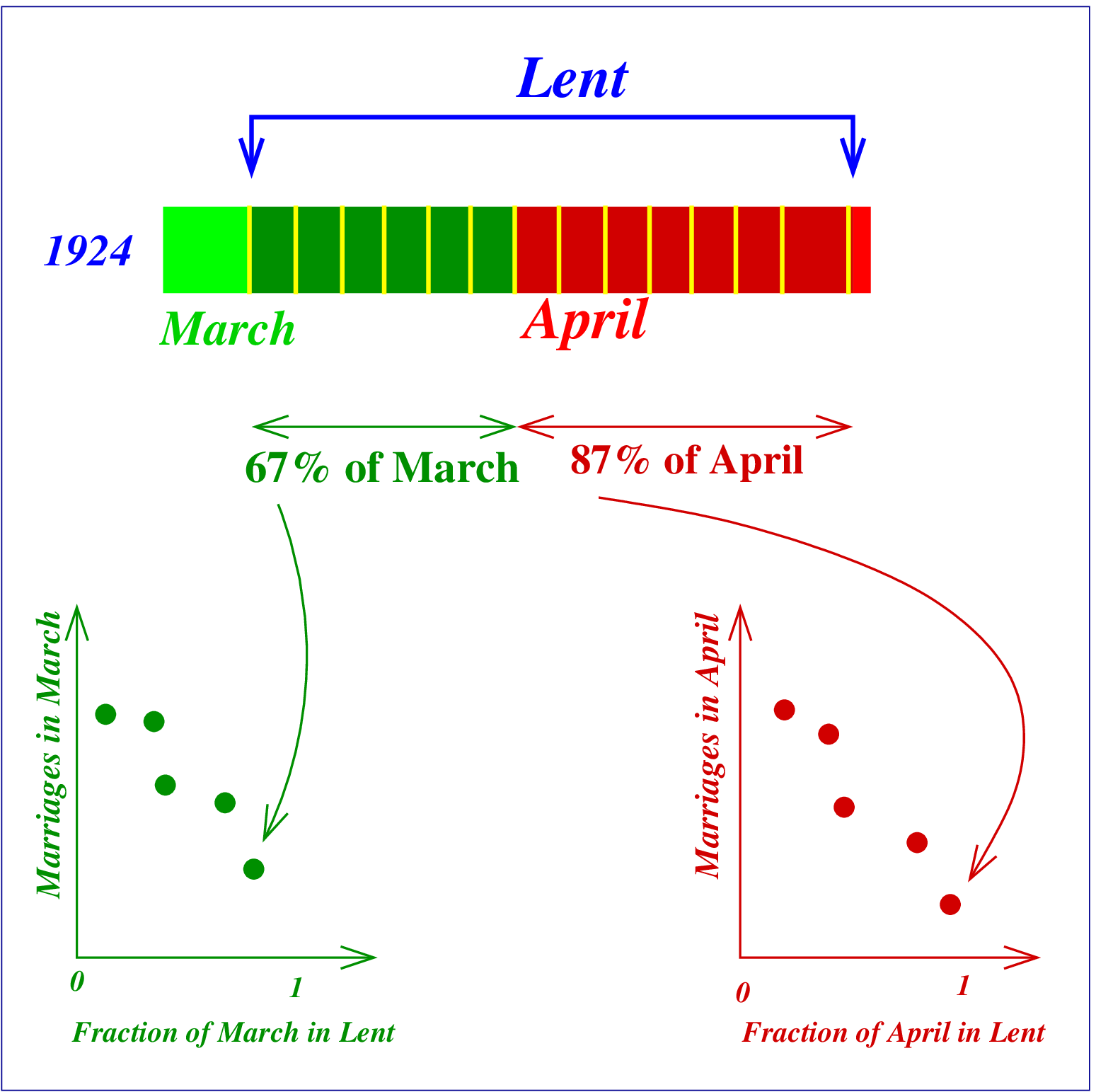}}
\qleg{Fig.\qhu 2\qhv Procedure for determining the correlation
between Lent-month overlap and the number of marriages in that
month.}
{In 1924 the Orthodox Lent lasted from 10 March to 26 April. 
The figure shows only March and April. In a few
years Lent may start in February which means that there
will be a third graph. It is important to keep the graphs of the
different months apart in order to avoid any interference
caused by the seasonal profile of marriages.}
{}
\end{figure}
%-------------------------------------------------
 
%
%%-----------------------------------------------
%%%%   COURBES MARIAGES ET INTERSECTION PR MARS + GRAPHE REGRESSION
\begin{figure}[htb]
\centerline{\psfig{width=17cm,figure=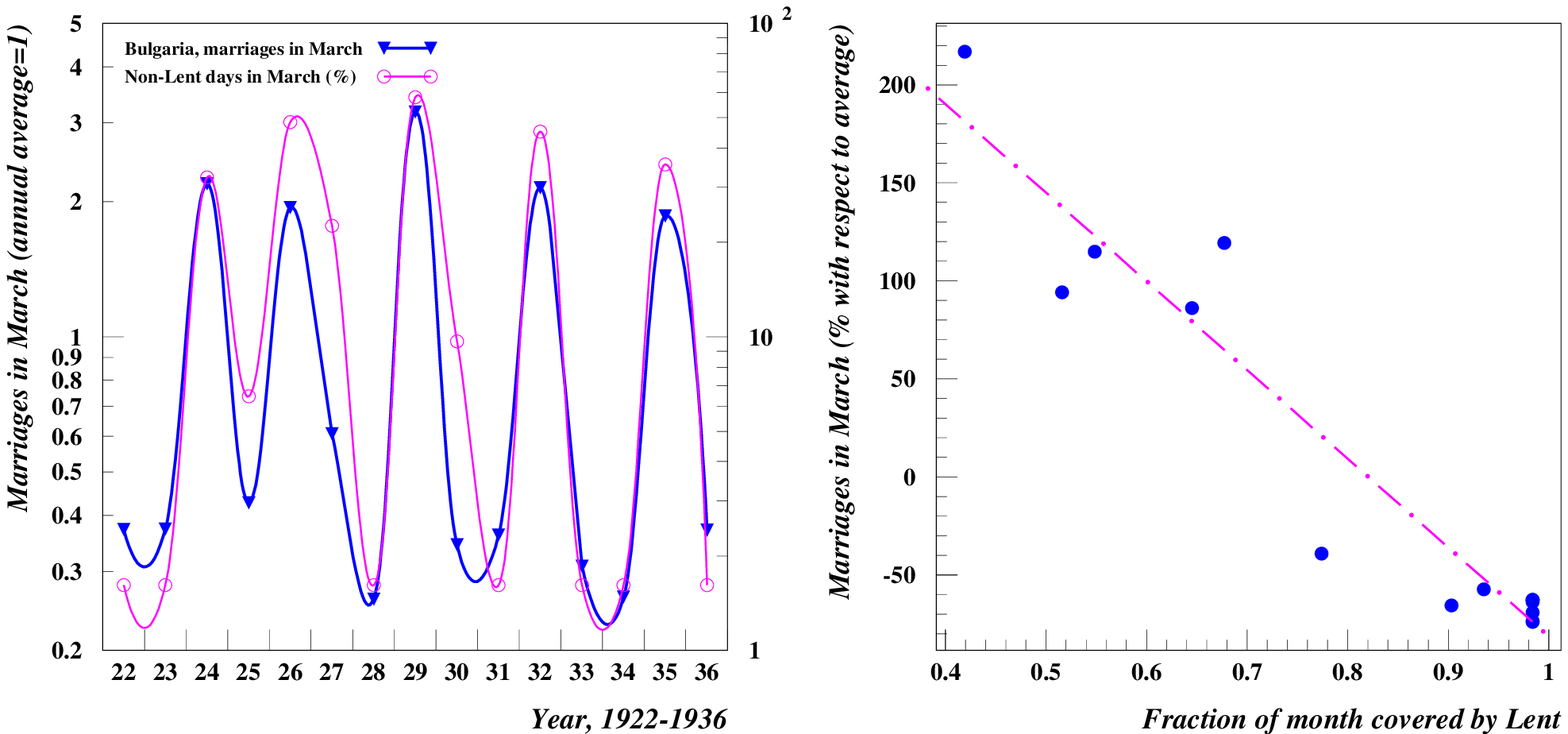}}
\qleg{Fig.\qhu 3a,b\qhv Number of marriages during March
in Bulgaria versus Lent in March.}
{{\bf (a)} Marriages versus non-Lent days. 
The scale for the non-Lent days is on the right-hand side.
{\bf (b)} Scatter-plot for the same data: $ x $=Lent days,
$ y $=marriages.
The correlation of the two series is 0.96.}
{Source: Bunle (1954, p. 249.}
\end{figure}
%-------------------------------------------------

Whereas for Bulgaria one gets a high correlation,
we can now also understand why the curve of the
marriages in France was out of sink. This is because,
except in a few years, the western Lent (i.e. Catholic
or Protestant) does not coincide with the Orthodox
Lent. The former starts earlier and is somewhat shorter.
It lasts 46 days instead of 49%
\qfoot{The western Lent lasts from ``Ashes Wednesday'' to
the Saturday before Easter. Actually the time for fasting
lasts only 40 days because the 6 Sundays are excluded.
Because, nonetheless all 46 days are a time
of penance, we treated them in a uniform way.
The same remark applies to the three days before Easter
which in some sources are not included
in Lent.}%
.
\qpar

The same procedure can be repeated for all months
with which Lent overlaps, that is to say February,
March and April. The fact that we get separate estimates
for each month is a distinct benefit of this procedure
for it permits to completely bypass the nagging difficulty
of how to get rid of the seasonal fluctuations of vital rates%
\qfoot{Ad hoc corrections can be tentatively defined
in various ways but lack clear theoretical
justification; moreover there is always the suspicion
that the observed effect has in some way been created
(or at least modified) by such corrections.}%
.     

\qA{Reduction in marriage numbers during Lent in Orthodox countries}

In Orthodox countries the clergy is very reluctant
to celebrate marriages during Lent. Thus, if all
marriages were celebrated in a religious way the
number of marriages during Lent would fall almost
to zero. As shown in
Fig. 3 and Table 1a, a sizeable number of 
marriages nonetheless take place during Lent;
this gives an estimate of the number of purely
civilian marriages 
\qpar

Table 1a gives the correlation and regression results for
Bulgaria, Romania and Greece. Lent has a drastic
effect on the number
of marriages: on average they are nearly divided by 4
and by 10 in the most extreme cases such as March 1928, 1929 
considered previously.
%
%%-----------------------------------------------
\begin{table}[htb]

\small
\centerline{\bf Table 1: Reduction in the number of marriages
during Lent in Orthodox countries}

\vskip 5mm
\hrule
\vskip 0.7mm
\hrule
\vskip 0.5mm
$$ \matrix{
\qtb
 & \hbox{March} & \hbox{April} & \hbox{Mar-Apr}\cr
 & \hbox{} & \hbox{} & \hbox{average}\cr
\noalign{\hrule}
\qth
\hbox{Bulgaria, 1922--1936} \hfill & & &\cr
\quad \hbox{Correlation } (l,m) \hfill& -0.96 & -0.88 & \color{red} -0.92\cr
\quad \hbox{Reduction, \%} (h) \hfill& -452\pm 74 & -177\pm 51 &
\color{red}-314\pm 44 \cr
\hbox{Romania, 1930--1936} \hfill& & &\cr
\quad \hbox{Correlation } (l,m\hfill) &-0.98 & -0.91 & \color{red} -0.95\cr
\quad \hbox{Reduction, \%} (h) \hfill&  -291\pm 38 & -70\pm 28 &
\color{red} -180 \pm 23 \cr
\hbox{Greece, 1931--1936} \hfill& & &\cr
\quad \hbox{Correlation } (l,m) \hfill& -0.99 & -0.99 & \color{red} -0.99\cr
\quad \hbox{Reduction, \%} (h)\hfill & -443\pm57 &-184\pm 19  &
\color{red} -313 \pm 27  \cr
\hbox{Average} \hfill& & &\cr
\quad \hbox{Correlation } (l,m)\hfill & & & \color{blue} -0.95\cr
\quad \hbox{Reduction, \%} (h) \hfill&  &  &
\color{blue}  -269\pm 22\cr
\noalign{\hrule}
} $$
\vskip 1.5mm
Notes: In the correlation $ m $ denotes the number of marriages
whereas $ l $ designates the overlap fraction between Lent
and the month under consideration. The reduction $ h $ is the
decrease in marriage number for a month completely
included in Lent, i.e. when the overlap fraction increases from 0 to
1. The error bars are confidence intervals for a confidence
probability of 0.95.
On average, during Lent the reduction is -269\% which means that
the number of marriages is divided by 3.7.
The results for February were omitted because the rare overlaps
produce too few data points.
\qL
Source: Bunle (1954, p. 248-251).
\vskip 2mm
\hrule
\vskip 0.7mm
\hrule
\end{table}
%%----------------------------------------------- 

\qA{Marriage reduction during Apostles' fast in Orthodox countries}

In Orthodox countries, apart from Lent and Advent there
is also the Apostles' Fast (also called ``Peter and Paul
Fast''). In a sense this is an ideal case for applying the
overlap methodology explained in Fig. 2 for this
fast falls always in June and is of variable length.
It starts on a date which can move from early
to late June and ends always on 29 June.
\qpar

If one applies the overlap procedure for 
Bulgaria in 1921-1936 one obtains for the
correlation between fraction
of June covered by fast and marriage rate a value  
of $ -0.39 $ (instead of $ -0.96 $ for Lent)
which shows that this fast is enforced with
much more flexibility both by the Orthodox Church and by the
population. As a matter of fact, in some denominations there
has been a secular tendency to reduce this fast to a few days instead
of its traditional length.  
\qpar

The same observation probably also
applies to Advent. However, in this case, since it is not a mobile
time interval one cannot apply the overlap methodology.

\qA{Reduction (or increase) in marriage number in Catholic and
Protestant countries.}

Table 1b documents the effect of Lent on marriages in two
Catholic countries namely France and Spain, one mixed Catholic-Protestant
country, namely the Netherlands and two Protestant countries,
namely Finland and Sweden.

%
%%-----------------------------------------------
\begin{table}[htb]

\small
\centerline{\bf Table 1b: Reduction in the number of marriages
during Lent in Catholic and Protestant countries}

\vskip 5mm
\hrule
\vskip 0.7mm
\hrule
\vskip 0.5mm
$$ \matrix{
\qtb
 &   \hbox{February}&\hbox{March} & \hbox{April} & \hbox{Feb-Apr}\cr
\qtb
 & \hbox{} & \hbox{} & &\hbox{average}\cr
\noalign{\hrule}
\qth
\hbox{France, 1922--1936} \hfill & & &&\cr
\quad \hbox{Correlation } (l,m) \hfill& -0.92& -0.63& -0.85& 
\color{red} -0.80\cr
\quad \hbox{Reduction, \%}\ (h) \hfill& -46\pm 15 & -90\pm 59 & -25\pm 8&
\color{red}-53\pm 16 \cr
\hbox{Netherlands, 1922--1936} \hfill& & &&\cr
\quad \hbox{Correlation } (l,m)\hfill & -0.88& -0.083 &-0.70& 
\color{red} -0.55\cr
\quad \hbox{Reduction, \%}\ (h) \hfill&  39\pm 15& 6\pm 41&-22\pm 12 &
\color{red} -22 \pm 13 \cr
\hbox{Spain, 1922--1936} \hfill& & &&\cr
\quad \hbox{Correlation } (l,m) \hfill&  -0.45& -0.57 & -0.77&
\color{red} -0.60\cr
\quad \hbox{Reduction, \%}\ (h)\hfill & 28\pm 41& -62\pm 49  & -31\pm 13&
\color{red} 40 \pm 20  \cr
\hbox{Average} \hfill& & & &\cr
\quad \hbox{Correlation } (l,m)\hfill & & & &\color{magenta}-0.65\cr
\quad \hbox{Reduction, \%}\ (h) \hfill&  &  & &
\color{magenta}  -38\pm 9 \cr
\hbox{Finland, 1922--1936} \hfill & & &&\cr
\quad \hbox{Correlation } (l,m) \hfill& -0.83& -0.55& +0.43& 
\color{red} \cr
\quad \hbox{Reduction, \%}\ (h) \hfill& -34\pm 17 & -44\pm 37& +15\pm 18&
\color{red}  \cr
\hbox{Sweden, 1922--1936} \hfill & & &&\cr
\quad \hbox{Correlation } (l,m) \hfill& -0.76& -0.25& +0.77& 
\color{red} \cr
\qtb
\quad \hbox{Reduction, \%}\ (h) \hfill& -29\pm 18 & -27\pm 58& +40\pm 18&
\color{red}  \cr
\noalign{\hrule}
} $$
\vskip 1.5mm
Notes: The meanings of $ l,m,h $ are the same as in Table 1a.
In France, the Netherlands and Spain,
on average, the reduction due to Lent is: -38\%.
In Finland and Sweden during the month of April instead of a
reduction one observes
an increase in the number of marriages
together with Lent overlap. The reason of this effect remains
an open question. For the sake of clarity only averages which
had a clear meaning were given in the table.
\qL
Source: Same as for Table 1a.
\vskip 2mm
\hrule
\vskip 0.7mm
\hrule
\end{table}
%%-----------------------------------------------

Two main observations emerge.
\qbu The impact of Lent on marriages 
is much smaller than
in Orthodox countries; in term of percentage it is seven times
smaller: 38\% as compared to 269\%.
\qbu Whereas in the great majority of cases Lent reduces marriage
numbers there are two cases where greater Lent overlap 
results in  {\it higher} marriage numbers. This happened during
the month of April in Finland and Sweden%
\qfoot{For Denmark there are no data available for 1931-1936.
Other Protestant countries which could be tested (and 
for which data are available) are Australia, Norway,
Scotland and the State of Massachusetts in the United States.
This will be left for a subsequent study.}%
.
The reason remains an open question.

\qI{Effect of Lent on conceptions}

\qA{Effect of new marriages on birth rates}

% Ce pt est discute ds MAPX#MARINAIS2 (PRELIMINAIRES)

Apart from new marriages does Lent also affect births?
Newly married couples may
conceive a child shortly after being married; in that way, 
fluctuations in the number of marriages might induce similar
fluctuations 9 months later in birth numbers.
Although this argument sounds plausible, a closer
examination will tell us that 
this effect is fairly small. \qL
The argument goes as follows.
\qparr

First we will describe three observations after which 
an explanation will be proposed.
\qpar

For the observations
we consider the series of monthly marriages, $ M(t) $,
and births, $ B(t) $, between the years $ t_1 $ and $ t_2 $.
Then, we translate $ B(t) $
to the left by 9 months which gives $ B'(t) $.
As this operation brings births into coincidence with conceptions
it should result in a substantial correlation.
This test will be carried out for three countries:
Bulgaria, Sweden and France.

\qun{Bulgaria} Here we take $ t_1=1930,t_2=1936 $ which
gives 84 monthly data points. The correlation between $ M(t) $
and $ B'(t) $ is found equal to $ 0.023 $ which means no correlation.

\qun{Sweden} The time interval is the same and leads to
a correlation of $ 0.30 $; the confidence interval with probability
0.95 is  (0.07,0.49) which means a low and barely significant
correlation.

\qun{France} Here we take $ t_1=1925,t_2=1936 $. The correlation
is found equal to $ 0.49 $ with a confidence interval
$ (0.34,0.60) $ which indicates a low but significant correlation. 

In short, in Bulgaria we found no correlation whereas in France
and Sweden we found a low correlation. 

\qA{Newly married women versus all women of child bearing age}

How can one
explain the previous results. Clearly they must be in relation with
the share of the conceptions which occur shortly after marriage
(itself a subset of first births) in the total of all births.
If we assume the same conception rate for all women
between the ages of 20 and 30 the number of their conceptions
will be  proportional to their numbers.
\qpar

We take the case of Sweden and for the sake of simplicity we 
limit our discussion to the age interval 20-30.
In the 1930s there were  on average
every month about 4,000 marriages which means
4,000 women joining the pool of all women able to have
children. At the census of 1930 there were 520,000 women
in the age group $ 20-29 $ (Flora et al. 1987, p. 72,134).
Thus, the $ 4,000 $ newly married women represented
only $ 0.8\% $ of the total number of women in the age-group
$ 20-29 $. This percentage will be even smaller in a country
like Bulgaria where the fertility rate is higher.
In 1930 in Sweden there were 56 births per 1,000 women
in the age interval $ 15-49 $ whereas there were 120 in
Bulgaria (Bunle 1954 p. 78-79).  
\qpar
In other words, it is not at all surprising that we
found no correlation in Bulgaria. It is rather the opposite
which is surprising, namely that we found a low correlation
in France and Sweden. This is probably due to the fact that the 
monthly fluctuations of births in the whole population are much
more stable than the monthly number of marriages.
\qpar

In the case of France which has the highest correlation we can
compute the regression. When each series is expressed in percentage
with respect to its average one gets: 
births=0.16$ \times $marriages.

\qA{Expected births versus observed births}

Can this coefficient of 0.16 by confirmed by
a direct calculation?
Fig.4 suggests an argument based on the sheer size of the
fall in birth rate as observed between October and December.
We focus on the case of Bulgaria.
For the interval 1930--1936 the average annual reduction in the
number of births was $ 6,180 $. We wish to compare this number to
the reduction expected as a consequence of the fall in marriages.
For the same 1930--1936 interval there were on average 13,100 marriages
in February compared with
only 1,900 in March. It results that the reduction
in the number of new married woman was $ 13,100-1,900=11,200 $.
As already mentioned above, there were 120 births per 1,000
women in the 15-49 age group. Thus, to the deficit of
$ 11,200 $ marriages will correspond an expected deficit of 
$ 11.2\times 120=1,340 $ births. How does this number compare
to the births reduction actually observed? As $ \rho=1,340/6,180=21.7\% $,
we see that this estimate of the share of the marriage effect
in the total birth reduction is compatible with the proportion
of $ 0.16 $ given by the regression.    
\qpar

The same calculation performed for France (also over 1930--1936)
leads to the following results.\qL
Annual reduction in births between October and November%
\qfoot{We took the difference Oct-Nov instead of Oct-Dec as previously
because from Nov to Dec there is an increase; this difference is
probably related to the fact that on average western Lent starts
about 2 weeks before Orthodox Lent.}%  
: $ 2,470 $ \qL
Annual reduction in marriages between February and March: $ 7,860 $ \qL
Number of births for 1,000 women aged 15-49: $ 68.2 $ \qL
Expected reduction in births due to the fall in marriages: 
$ 7.86\times 68.2 = 535 $ \qL
Ratio of expected to observed reduction in births: $ \rho=535/2470=21.6\% $
\qpar
The fact that the two values of $ \rho $ are close to each other 
in spite of a difference in the magnitude of the effect is reassuring,
\qpar 

In conclusion 
we now know that the fall in marriages accounts for not more
than 25\% of the observed decrease in birth numbers. The 75\%
remaining are certainly due to a reduction
in conceptions that is to say a change in sexual behavior.
Next we present two
graphs which give an intuitive  idea of this effect.
After that we will try to estimate the relationship between Lent 
and births. 

\qA{Evidence for the effect of Lent on births 9 months later}

In the first section it was suggested that the origin of
the fluctuations of marriage numbers in Bulgaria could be
identified through comparative analysis.
The same message could be repeated here for indeed if considered
alone the curves of Fig. 4a,b would not be easy to explain.
However, a comparison with Japan and France would again suggest that
this effect is related with the Orthodox religion.
 
%
%%-----------------------------------------------
%%%%   NAISSANCES 
\begin{figure}[htb]
\centerline{\psfig{width=9cm,figure=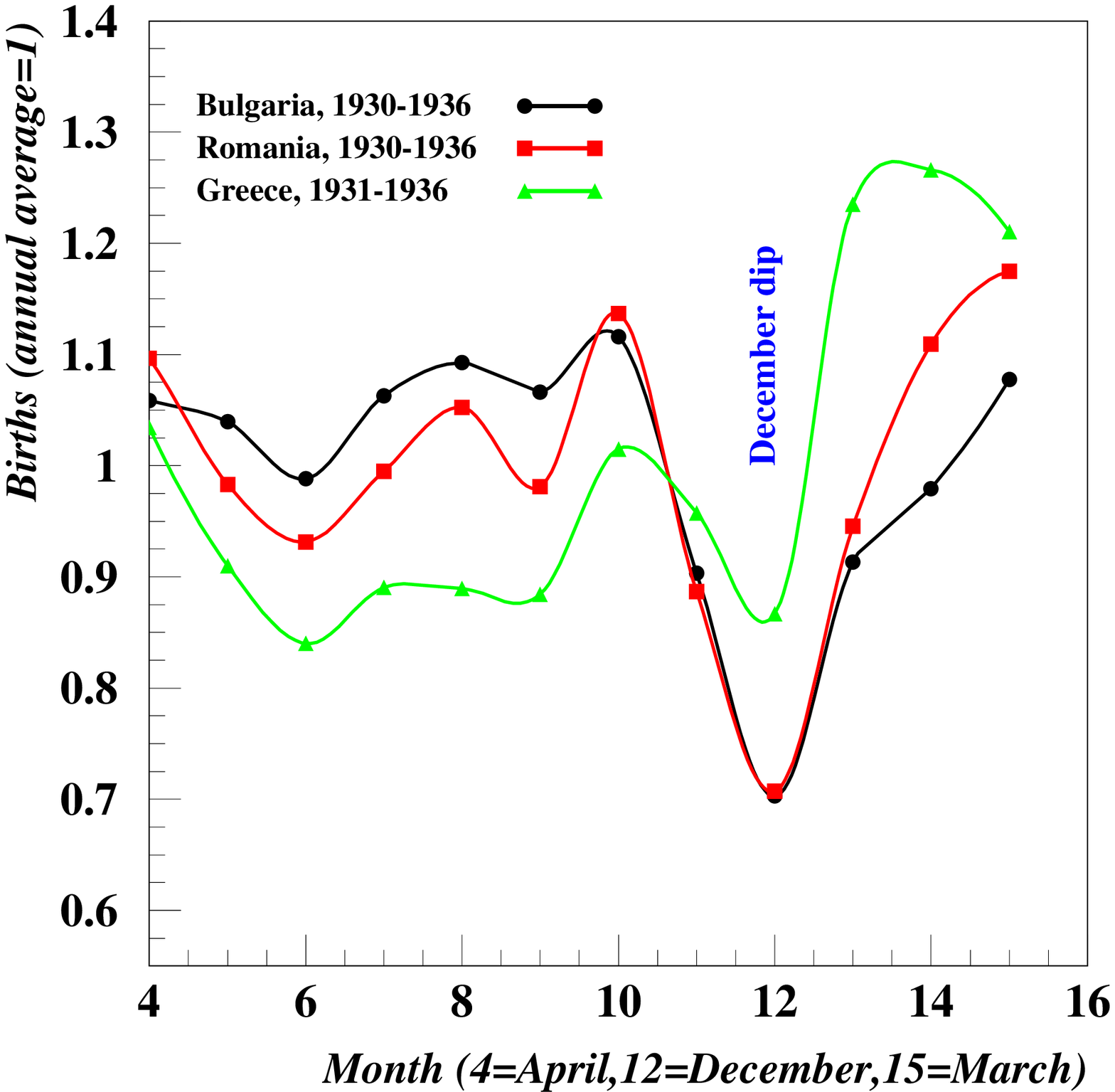}}
\qleg{Fig.\qhu 4a\qhv Number of births in Bulgaria from April to March
of the following year.}
{In the context of this paper it will be clear to readers
that the dip is due to the effect of Lent 9 months earlier.
However, considered alone it would appear fairly mysterious.}
{Source: Bunle (1954, p. 307,309,311.}
\end{figure}
%-------------------------------------------------

%
%%-----------------------------------------------
%%%%   COURBES NAISSANCES ET INTERSECTION PR AVRIL->JAN
\begin{figure}[htb]
\centerline{\psfig{width=9cm,figure=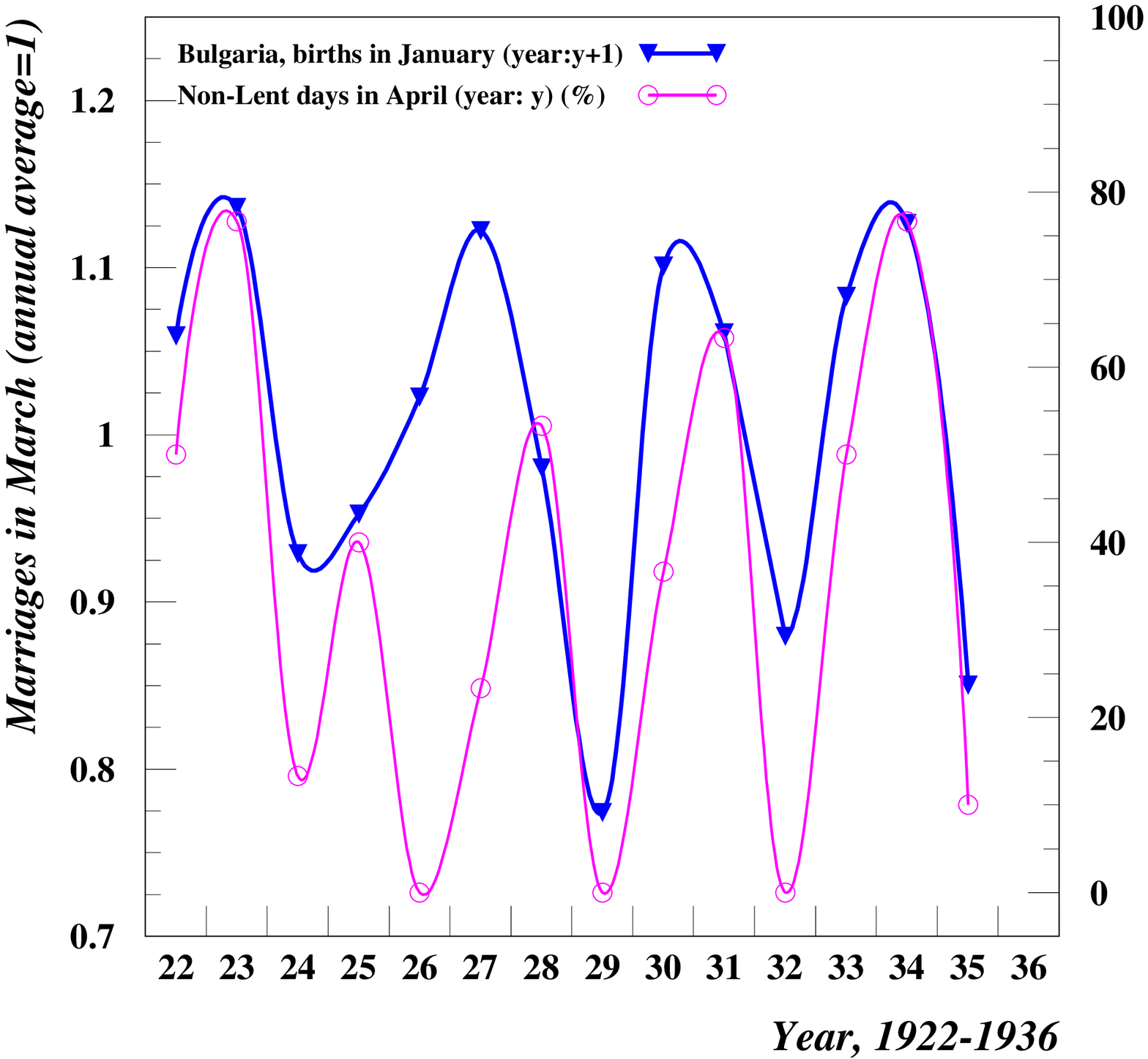}}
\qleg{Fig.\qhu 4b\qhv Number of births during January
in Bulgaria versus number of non-Lent days in April
of the previous year.}
{The scale for the non-Lent days is on the right-hand side.
The correlation of the two series is 0.72.
This figure parallels Fig.3a
which showed the correlation between marriages and non-Lent days.
However, in order for the conceptions of April to be mapped
into births occurring in January of the following year,
the pregnancy length (from conception to birth)
has to be exactly equal to 9 months (i.e.
273 days). If its real average length is different (e.g. 280 days)
the synchronicity between Lent and births will be affected.}
{Source: Bunle (1954, p.307).}
\end{figure}
%-------------------------------------------------

\qA{Two difficulties}

The effect of Lent on births is more difficult to study than
its impact on marriages for two reasons.
\qbu By comparing Fig.3a and 4b one sees that for births the
effect is much smaller than for marriages.
This implies a greater relative incidence of background noise.
Thus, one should not be surprised to see larger error bars. 
\qbu The length of time between sexual intercourse and birth
is not known exactly
In medical practice
it is customary to define the length of pregnancy
as the interval between the first day of the woman's last period
and birth. Conception usually occurs two weeks later, however.
If one accept the traditional figure of
40 weeks (i.e. 280 days) for pregnancy in the medical sense
it results in a
conception-birth interval of 266 days. Subsequently,
we will use the word ``pregnancy'' as referring to
the length time between sexual intercourse (which is zero
one or at most two days before conception) and birth;
it will be denoted by $ G $. 
Fig. 4b was drawn on the assumption that this time interval
was 9 months (i.e. 273 days) but this is only an
approximation. The analysis presented in the next subsection
will allow us to determine the  length of pregnancy
that is ``statistically optimum''.

\qA{Analysis of the relationship between Lent and births}

In this study a simple but important tool
was a computer macro which for any
day defined by its date (year=$ y $, month=$ m $ and day
of month=$ j $) computes its position index $ i $  in number of
days with respect to a given origin, for instance 1 January 1920.
For each of the three months November, December and January the
calculation was conducted in the following way
(the explanations are for November 1920).
\qee{1} First, the Lent interval of 1920 was translated by $ G $ days
to the right to an interval $ (i_1,i_2) $ 
\qee{2} Secondly, the overlap of $ (i_1,i_2) $ with November
was determined. We call it $ L $.
\qee{3} The number of birth $ m $ in November 1920
was taken from Bunle (1954) 
\qee{4} Once the pairs $ (L,m) $ have been computed
for each of the years under consideration
(e.g. 1920-1936) the correlation $ r $ and
the regression of the $ (L,m) $ scatter-plot were computed. 
From the regression slope
one derives the change of $ m $ when $ L $
changes from 0 to 1. Once expressed in percent, this change is
denoted by $ h $. As for higher $ L $ it is a fall in birth numbers 
that is expected, $ r $ and $ h $
should be negative; therefore it is natural to consider as 
optimum the values of $ G $ which give the most negative 
correlations.
\qee{5} As in the procedure used for marriages
the months of November,
December and January are treated separately. Thus, our results are
independent of the seasonal distribution of births. 
As already emphasized this is a crucial advantage.
However,
if one wishes to get synthetic indicators one can take the
averages $ \overline r $ and $ \overline h $  of the three values 
$ r_1,r_2,r_3 $ and  $ h_1,h_2,h_3 $. 
An additional benefit of averaging
is to reduce the error bars by a factor $ \sqrt{3} $.
\qpar

\qun{Testing the procedure}
In order to test the previous procedure a simulated birth
series was built. We started with a constant
daily series of birth numbers; then, in each year, 
for the days belonging to Lent
plus $ G $ days
(with real dates for Lent) the birth numbers were reduced by 
$ q $ percent. After that, the daily series was converted
into a monthly series.
Finally,
a variable amount of random noise could be added to make the
series look more realistic. However it is with a purely
deterministic series (i.e. without noise) that the most interesting
tests can be done. In this case one should get $ r_1=r_2=r_3=1 $
and $ h_1=h_2=h_3=q $ which is indeed what was verified.
\qpar

Another instructive test is with respect to the value of $ G $.
This point is
explained in the next subsection.

\qA{Influence of pregnancy length and its significance}
Fig. 5a,b shows that the value of the assumed pregnancy
length $ G $ used in the analysis strongly affects the
results. In order to assess the significance of these
results it is useful to carry out a test 
with a simulated series. 
\qpar

In this test
we built the simulated series
with a pregnancy length $ G_1 $ and we analyze it with a different
value $ G_2 $. This test shows that even a small
difference between $ G_1 $ and $ G_2 $ may lead to greatly
reduced correlations. For instance, with $ G_1=273 $ days and
$ G_2=276 $, the value of $ r_2 $ falls from 1 to about 0.5
whereas $ r_1 $ and $ r_3 $ are hardly affected. This difference
between $ r_2 $ on the one hand and $ r_1,r_3 $ on the other
hand is of course related to the positions of Lent+$ G $.
Because in most years a large part of December (or even the
whole of it) is included in Lent+$ G $, most of the $ L $ values
of this month
are close to 1. This makes the scatter plot of December fairly unstable
which explains that even a small perturbation may change it
completely.
\qpar

%
%%-----------------------------------------------
%%%%   COURBES MONTRANT L'INFLUENCE DE G
\begin{figure}[htb]
\centerline{\psfig{width=17cm,figure=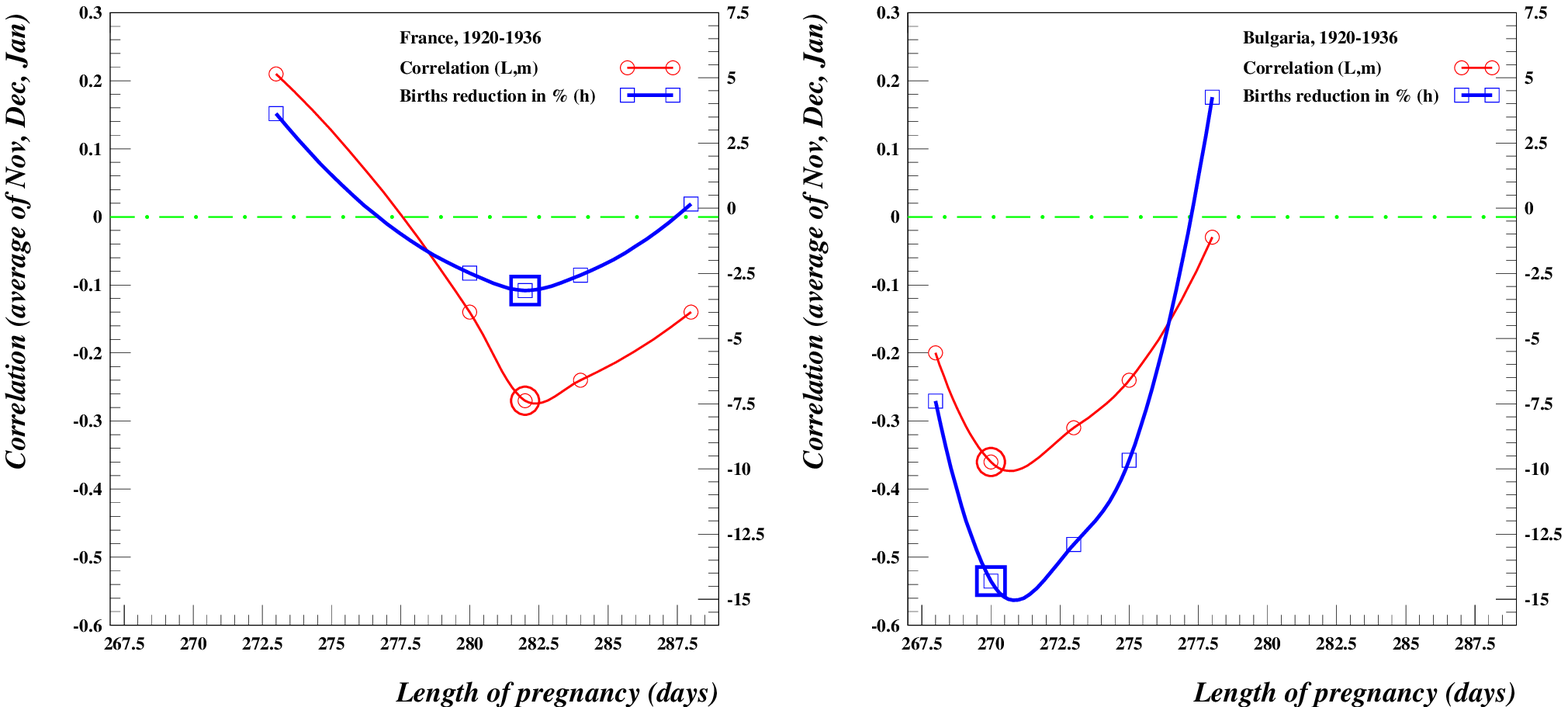}}
\qleg{Fig.\qhu 5a,b\qhv Relationship between Lent and number of births
as a function of assumed pregnancy length ($ G $).}
{In the correlation $ (L,m) $, the variable
$ L $ represents the overlap of Lent+$ G $
with the month under consideration;
$ m $ represents the number of births in the same month.
The graph shows the average correlation for the three months
of November, December and January (of the next year).
In the reduction $ h $ represents the birth reduction which
occurs when the month is completely included in Lent
with respect to the case when Lent does not overlap with 
this month. The comparison of {\bf (a)} and {\bf (b)} shows
that whereas the correlation is approximately the same in France
and Bulgaria the effect of Lent on births is about three times
larger in Bulgaria. The fact that the optimum values of $ G $ 
are not the same may be due to the fact that in some places 
it is the registration day which was recorded rather than the real
birth day. If this reason can be excluded, the present observation
can be related to the fact that the natural
variation in pregnancy length is fairly large, in fact 
larger than thought previously
(ScienceDaily 2013). It can be noted that
Romania and Greece the optimum $ G $ value
is also around 272 days.}
{}
\end{figure}
%-------------------------------------------------

This test suggests that the differences documented in 
Fig. 5a,b are of significance and that the value of $ G $
which gives the highest correlation identifies the $ G $ of
the real world. In order to substantiate this judgment
it would be helpful to find accurate estimates of pregnancy
length in different countries. An Internet search was
unsuccessful and this is in fact not surprising. 
Accurate measurement
requires exact determination of the moment of ovulation
which in turn relies on daily blood tests for hormone levels.
As this is a demanding procedure it is understandable that
it is not done frequently. Another difficulty comes from
the fact that whenever dedicated campaigns were set up
they revealed larger standard deviations
than expected (Bhat et al. 2006, ScienceDaily 2013).
Within a given country the standard deviation is about
7 days but this does not tell us what is the cross-country 
variability.

\qA{Strength of the Lent effect in Orthodox countries}

Fig. 5a,b shows the Lent effect is about three times smaller
in a western country like France than in an Orthodox country
like Bulgaria. The fact that the effect is smaller also means that
it is more difficult to measure. That is why here we will mostly focus
on Orthodox countries. 
%
%%-----------------------------------------------
\begin{table}[htb]

\small
\centerline{\bf Table 2a: Reduction in the number of conceptions
during Lent in Orthodox countries, 1930-1936}

\vskip 5mm
\hrule
\vskip 0.7mm
\hrule
\vskip 0.5mm
$$ \matrix{
\qtb
 &   \hbox{Bulgaria}&\hbox{Greece} & \hbox{Romania} \cr
\noalign{\hrule}
\qth
\hbox{(1) Reduction }h\hbox{ expressed in \%},\hbox{ Nov,Jan}\hfill
& -32\pm 22 & -44\pm 11&-16\pm 15\cr
\qtb
\hbox{(2) Reduction }h\hbox{ expressed in \%},\hbox{ Nov,Dec,Jan}\hfill
& -13\pm 18 & -27\pm 8& -9.5\pm 11\cr
\noalign{\hrule}
} $$
\vskip 1.5mm
Notes: $ h $ is the reduction 
in birth number (expressed in percent with respect to the
situation without Lent)
when a given month is completely included in Lent.
As explained in the text the December estimates are very
uncertain because most $ L $ values are close to 1; here
all December estimates in fact turn out to be positive. 
That is why we give the results in two forms: in (1) the average
is restricted to Nov and Jan whereas in (2) the average
includes all three months. Naturally, the estimates of (2) are lower
but the ranking of the countries is not changed. Note that for
Greece the time interval is 1931--1936 because the data for 1930
are not available. Finally, it should be recalled that about
1/4 of this effect is the consequence of a fall in marriages
9 months earlier whereas the remaining 3/4 results from a drop
in conceptions during Lent. 
\qL
Source: Bunle (1954,p.307-311) 
\vskip 2mm
\hrule
\vskip 0.7mm
\hrule
\end{table}
%%-----------------------------------------------
%
%
%%-----------------------------------------------
\begin{table}[htb]

\small
\centerline{\bf Table 2b: Reduction in the number of conceptions
during Lent in France}

\vskip 5mm
\hrule
\vskip 0.7mm
\hrule
\vskip 0.5mm
$$ \matrix{
\qtb
 &   1872-1891 & 1920-1936 \cr
\noalign{\hrule}
\qth
\hbox{(1) Reduction }h\hbox{ expressed in \%},\hbox{ Nov,Jan}\hfill
&  2.6\pm 2.5 & -2.0\pm 4\cr
\qtb
\hbox{(2) Reduction }h\hbox{ expressed in \%},\hbox{ Nov,Dec,Jan}\hfill
& -0.29\pm 4.6 & -2.9\pm 3 \cr
\noalign{\hrule}
} $$
\vskip 1.5mm
Notes: The comments already made in Table 2a also apply here.
As in Table 2a and for the same reason the inclusion of December
whose regression slope is quite unstable (sometimes negative,
sometimes positive)
makes the results given in line 2 fairly uncertain. 
As the effect of Lent is about 5 to 10 times weaker than in
Orthodox countries, the error bars are much larger in relative terms.
Confidence in the fact that the estimates of $ h $ are fairly
correct despite the large error bars comes from the 
observation that 
similar results are obtained for different values of $ G $.
\qL
Sources: The monthly birth data for 1872--1891 are from
``Statistique G\'en\'erale de la France'' (several years).
The data for 1920-1936 are from Bunle (1954,p.308),  
\vskip 2mm
\hrule
\vskip 0.7mm
\hrule
\end{table}
%%-----------------------------------------------
%
to the incidence of religion on other vital rates.
Table 2  summarizes the results for
Bulgaria, Greece and Romania.
It appears that the strength
of the Lent effect is strongest in Greece.

\qI{Conclusion}

First we have studied the impact of Lent on marriage rates.
This was fairly easy because there is a very strong connection.
It is somewhat stronger in Orthodox countries than in Western
countries but is clearly visible in both cases. 
\qpar

The effect of Lent on conceptions was more difficult to study
but it was also more interesting. Why? \qL
Whether we consider religious or civil marriages, they will
be subject to the scrutiny of the social group
in which they take place. In other words, they are an indicator
of social consensus and conformity. Religious consensus
can be measured through other indicators. One can mention
the following; the data for France are given
for  the purpose of illustration; they are from 
``Statistiques de l'Eglise catholique en France'' (2017).
\qee{1} The proportion of the population that is baptized.
Being a global indicator, this proportion will change very
slowly in the course of time. A more revealing indicator 
is the proportion of baptized person by age group.
In 2015 in France 33\% of newborns were baptized,
down from 61\% in 1990.
\qee{2} The proportion of religious marriages. 
Clearly, this indicator
is only meaningful in countries where the notion of purely
civil marriage is recognized%
\qfoot{According to the Wikipedia article entitled
``Civil marriage'' the set of countries where 
it exists only a religious
marriage comprise the following:
Bahrain,
Egypt,  Indonesia, Iran, Israel, Jordan, Lebanon, Libya, 
Qatar, Syria, Saudi Arabia, United Arab Emirates, Yemen.}%
.
In 2015 in France, the percentage of Catholic marriages
was 24\%, down from 51\% in 1990.
\qee{3} The proportion of priests and nuns in the population.
In 2012 in France priests and nuns represented 0.74 per million
population, down from 1.30 in 2000.
\qee{4} The proportion of the population who attend weekly religious
service. No French data were available for this indicator.
\qpar
It can be noted that the rates of decline of all French indicators
are roughly the same: about 50\% in 25 years, i.e. 2\% per year. 
\qpar

With respect to the previous indicators, the incidence of religion
on conceptions is of a different kind in the sense that it is
independent of any form of social control. Therefore one is not
surprised that in terms of percentage the reduction
is about ten times smaller than the incidence on
marriages.
\qpar

In the future it will be possible to extend this exploration
to the incidence of mobile religious time periods
on other vital rates. There
are mobile time intervals similar to Lent in other religions,
e.g. the ``Ten Days of Repentance'' 
(from Rosh Hashanah to Yom Kippur in September-October)
in the Jewish religion or Ramadan in Islam.
Some preliminary tests have revealed that there is a decrease in
suicide rates during the month of Ramadan. Is there a similar decrease
during Lent or during the ``Ten Days of Repentance''? 
Thanks to the methodology introduced in the present paper, such
issues can be investigated by using monthly instead of daily data.
In contrast to
daily data which are publicly available
only in a few countries, monthly data are commonly available.

\vskip 10mm

{\bf References}

\qparr
Bhat (R.A.), Kushtagi (P.) 2006: 
A re-look at the duration of human pregnancy.
Singapore Medical Journal 47,12,1044-1048.

\qparr
Bunle (H.) 1954: Le mouvement naturel de la population dans le
monde de 1906 \`a 1936. [Vital statistics of
many countries world-wide from 1906 to 1936.]
Editions de l'Institut d'Etudes D\'emographiques, Paris.

\qparr
Herteliu (C.), Ileanu (B.V.), Ausloos (M.), Rotundo (G.) 2015:
Effect of religious rules on time of conception in Romania
from 1905 to 2001.
Human Reproduction 30,9,2202-2214.

\qparr
Herteliu (C.), Richmond (P.), Roehner (B.M.) 2018: 
Deciphering the fluctuations of high frequency birth rates.
arXiv:1802.08966.

\qparr
ScienceDaily-Oxford University Press 2013: 
Length of human pregnancies can vary
naturally by as much as five weeks. 
ScienceDaily, 6 August 2013. 

\qparr
Statistiques de l'Eglise catholique en France 2017.
(available on Internet)

\end{document}